\documentclass[prb,twocolumn,showpacs]{revtex4}
\usepackage{amsmath}
\usepackage{amsfonts}
\usepackage{amssymb}
\usepackage{hyperref}
\usepackage{epsfig}
\usepackage{epstopdf}
\usepackage{graphicx}
\usepackage{enumerate}
\usepackage{multirow}
\usepackage{verbatim}

\begin{document}

\title{Theory of  oxygen $K$-edge x-ray absorption spectra of cuprates}
\author{Xin Wang$^1$, Luca de' Medici$^2$, and A. J. Millis$^1$}
\affiliation{$^1$Department of Physics, Columbia University, 538
West 120$^{th}$ Street, New York, New York 10027, USA\\
$^2$Laboratoire de Physique des Solides, Universit\'e Paris-Sud,
CNRS, UMR 8502, F-91405 Orsay Cedex, France}
\date{\today}

\begin{comment}
71.27.+a    Strongly correlated electron systems; heavy fermions

78.70.Dm    X-ray absorption spectra

71.30.+h Metal-insulator transitions and other electronic
transitions

74.25.-q    Properties of superconductors \\
74.25.Jb    Electronic structure (photoemission, etc.)

74.72.-h    Cuprate superconductors\\
74.72.Gh    Hole-doped
\end{comment}

\begin{abstract}
The dynamical mean-field theory of the three-band model of
copper-oxide superconductors is used to calculate the doping
dependence of the intensity of the  oxygen $K$-edge x-ray absorption
spectra of high-$T_c$ copper-oxide superconductors. The model is
found not to reproduce the results of a recent experiment,
suggesting that at sufficiently high doping the physics  beyond the
conventional three-band model becomes important.
\end{abstract}
\maketitle

\section{Introduction}

The important structural unit of  the high temperature copper-oxide
superconductors is the ``CuO$_2$ plane'' and the important
electronic orbitals are believed to be the Cu $3d_{x^2-y^2}$ and O
$2p_\sigma$ states. High-$T_c$ superconductors are formed by doping
(adding carriers to) an insulating parent compound. The nominal
electronic configuration of the insulating parent compound is Cu
$3d^9$ (one hole in $d_{x^2-y^2}$ band) O $2p^6$ (oxygen $p$-band
completely full), although of course Cu-O hybridization means that
some fraction of the hole delocalizes onto the O. However, the
strong correlations characteristic of the Cu $d$ orbitals implies
that the energy required to add a second hole on the Cu site is very large,
$U\sim9$eV\cite{Mila88,Veenendal94} so that at low doping levels,
doped holes go primarily onto the oxygen site, but are strongly
coupled via an exchange interaction to the holes on the Cu sites,
forming ``Zhang-Rice singlets''.\cite{Zhang88} While this picture is
qualitative, being strictly valid only in the strong coupling limit
and low doping limits, it has provided a useful guide for thinking
about the materials. An important open question concerns the doping
at which this picture breaks down.

Oxygen $K$-edge x-ray absorption spectra provides an interesting
test of the Zhang-Rice picture, and more generally of our understanding
of the three-band model. In these experiments absorption of
an incident photon promotes an electron from the oxygen $1s$ shell
to an unoccupied oxygen orbital; if the photon energy is
appropriately tuned, the final states are in the energy range of the
O $2p$ manifold and the spectrum reveals the energy distribution of
the unoccupied O $2p$ states (modified by the excitonic
core-hole/excited state interactions). Experiments performed in the
early 1990s\cite{Chen91} on insulating and lightly-doped samples
La$_{2-x}$Sr$_x$CuO$_{4\pm\delta}$ (LSCO) revealed a two-peak
structure, with a higher energy peak visible in both insulating and
hole-doped materials and a lower energy peak, visible only in hole
doped compounds. The integrated spectral weight of the lower energy
feature was found to increase linearly with doping (at low hole-dopings) whereas the integrated spectral weight of the higher energy
feature was found to decrease.\cite{Chen91,Hybertsen92} For this
reason the higher energy peak was interpreted as the ``upper Hubbard
band'', while the lower energy feature was interpreted as the
``Zhang-Rice band''.

\begin{figure}
    \centering
    \includegraphics[width=0.75\columnwidth, angle=-90]{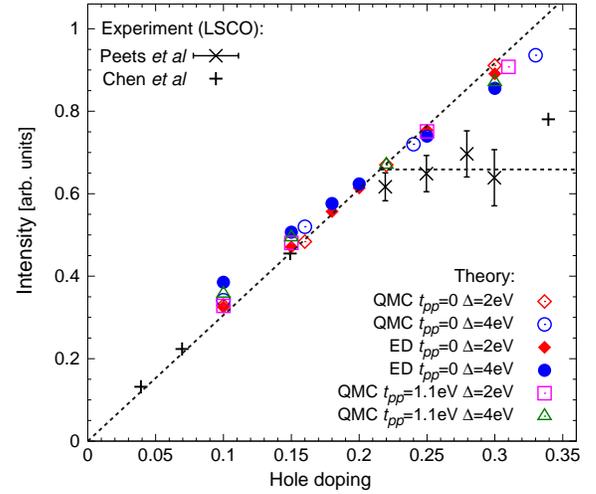}
    \caption{Relative intensity of the peak associated with the Zhang-Rice band of O $K$-edge x-ray
    absorption spectra v.s. doping. The plus-sign shows experiments on LSCO by Chen \emph{et al}
    (Ref.~\onlinecite{Chen91}) and the ``x''-shaped sign shows experiments on LSCO by Peets \emph{et al}
    (Ref.~\onlinecite{Peets09}). The diamonds, circles, squares and triangles are theoretically calculated
    integrated spectral weight of unoccupied oxygen states, normalized such that their values at 0.22 hole
    doping are the same.
    Dashed lines are only guides to eyes.
    Parameters: $U=9eV$, $t_{pd}=1.6eV$, $T=0.1eV$, $t_{pp}$
    and $\Delta=\varepsilon_p-\varepsilon_d$ as shown in the legends.}
    \label{ZR}
\end{figure}

A recent experiment\cite{Peets09} has extended the measurements to a
wider  range of compounds and in particular to a wider range of hole
dopings. Fig.~\ref{ZR} summarizes the data, showing the recent
measurements  as ``x''-shaped crosses with errorbars and the
previous  data\cite{Chen91} as plus signs.  A remarkable finding of
the recent measurements is that the oxygen weight in the feature
interpreted as the  Zhang-Rice band saturates as doping is increased
beyond the optimal doping value which maximizes the superconducting
transition temperature $T_c$. Ref.~\onlinecite{Peets09} interpreted
this finding as a breakdown of the Zhang-Rice singlet picture.

Motivated by these experiments, in this paper we use a more modern
theoretical technique, namely single-site dynamical mean-field
theory \cite{Georges96, Kotliar06} to compute the oxygen $K$-edge
x-ray absorption spectra implied by  the standard three-band
copper-oxygen model of high-$T_c$ cuprate superconductors. The model
contains the physics of Zhang-Rice singlets at low doping and strong correlations,
but for higher doping or weaker correlations the behavior becomes band-like.
We have examined the model both at strong coupling, where the state
at half filling is a charge-transfer insulator\cite{Zaanen85} and the physics at low doping is of Zhang-Rice
singlets, and at weaker correlation strength, where antiferromagnetism is needed
to make the model insulating at half filling and the appropriateness of the Zhang-Rice picture
is less clear.
We find that in both cases the
model reproduces the low doping behvior but qualitatively fails to reproduce the experimental
data at higher doping.\cite{Peets09} We conclude that if the experiment indeed measures
the oxygen density of states without other complicating effects, then the data indicate not just a failure of the
Zhang-Rice model but a breakdown of the entire three-band model
picture, with some additional orbital, not included in the three-band model, becoming important.

An additional consequence of our work is that by comparing the energies of the ``Zhang-Rice band'' and ``upper Hubbard band'' features to our calculations we are able to estimate the correlation strength of the materials.
The comparison places the materials on the metallic side of the metal/charge-transfer-insulater phase diagram, in agreement with previous work.\cite{demedici09,Comanac08}

The rest of this paper is organized as follows: Section \ref{model}
describes the three-band model to be studied and the methods. In
section \ref{result} we present the numerical results. Section
\ref{conclusion} is a conclusion and discussion.

\section{Model and Methods}\label{model}

The three-band Emery model considers Cu $3d_{x^2-y^2}$ and O
$2p_\sigma$ orbitals,\cite{Emery87,Varma87} which hybridize via a
copper-oxygen hopping $t_{pd}$.  We shall also consider
oxygen-oxygen hoppings, adopting the form implied by the
considerations of  Ref.~\onlinecite{Andersen95}.  The model exhibits
both a low-doping antiferromagnetic phase and a high-doping
paramagnetic phase but because the experimentally interesting
behavior occurs at high dopings, we restrict our attention in the
paramagnetic phase. We distinguish the oxygen sites displaced from
the Cu in the $x$ and $y$ directions, adopt the basis
$|\psi\rangle=\left({d_{k\sigma}}, {p_{x,k\sigma}},
{p_{y,k\sigma}}\right)$, introduce the copper-oxygen hopping
$t_{pd}=1.6eV$ and oxygen-oxygen hopping  $t_{pp}$  in the form
proposed by Andersen;\cite{Andersen95} we compare $t_{pp}=1.1eV$ and
$t_{pp}=0$.\cite{demedici09} The interaction part of the Hamiltonian
is
\begin{equation}
 H_{\rm int}=U\sum_in_{d\uparrow}n_{d\downarrow},\label{Hint}
\end{equation}
and in the calculation presented here we take $U=9eV$ (the precise value of $U$ is not important as long as it is larger than about $5eV$). The band theoretic part of the Hamiltonian may be represented as a $3\times3$ matrix:
\begin{widetext}
\begin{eqnarray}
{\bf H}_{\rm 3band}=
\left(\begin{array}{ccc}
\varepsilon_d & 2it_{pd}\sin\frac{k_x}{2} & 2it_{pd}\sin\frac{k_y}{2}\\
-2it_{pd}\sin\frac{k_x}{2} & \varepsilon_p+2t_{pp}(\cos k_x-1) & 4t_{pp}\sin\frac{k_x}{2}\sin\frac{k_y}{2}\\
-2it_{pd}\sin\frac{k_y}{2} &
4t_{pp}\sin\frac{k_x}{2}\sin\frac{k_y}{2} &
\varepsilon_p+2t_{pp}(\cos k_y-1)
\end{array}\right).
\end{eqnarray}
\end{widetext}

The model has two important parameters: $U$ and
$\Delta=\varepsilon_p-\varepsilon_d$. The physically relevant case
is large $U$, in which case the model is in the charge-transfer
regime \cite{Zaanen85} and the physics is controlled by the ratio of
$\Delta$ to the copper-oxygen hopping $t_{pd}$.   At ``half
filling'' (one hole in the $d$-$p$ complex)  the single-site
dynamical mean-field approximation predicts that  the model exhibits
a paramagnetic insulating phase for small $\Delta$  whereas the
paramagnetic phase is  metallic at large $\Delta$. In previous
work\cite{demedici09} we have determined that for $U=9eV$,
$\Delta=2eV$ is on the insulating side, but not far from the
metal-insulator phase boundary, whereas $\Delta=4eV$ lies on the
metallic side of the phase diagram, but also not far from the
boundary. The correct value of $\Delta$ for the cuprates is
controversial\cite{Comanac08}; we therefore consider $\Delta$-values
corresponding both to paramagnetic metal and to paramagnetic
insulating phases at half filling.

The model is solved using the single-site dynamical mean-field
approximation,\cite{Georges96,Kotliar06} in which the key
approximation is a momentum-independent self-energy:
\begin{equation}
{\mathbf \Sigma}(z, {\mathbf k})\rightarrow{\mathbf
\Sigma}(z)=\left(\begin{array}{ccc}
\Sigma(z) & 0 & 0\\
0 & 0 & 0\\
0 & 0 & 0\end{array}\right),
\end{equation}
where $z$ indicates real or Matsubara frequencies. The Green's function is
\begin{equation}
{\bf G}(z, {\mathbf k})=\left(z{\bf 1}+\mu-{\bf \Sigma}(z)-{\bf
H}_{\rm 3band}\right)^{-1},
\end{equation}
and the spectral functions (density of states) of $d$ and $p$
orbitals are obtained from the imaginary part of the diagonal
elements of the Green's function matrix.

We employ two impurity solvers: Exact Diagonalizations
(ED)\cite{Caffarel94,Capone04} and the hybridization-expansion
continuous-time quantum Monte Carlo method (QMC)\cite{Werner06}.
Because the two methods involve different approximations, comparison
of the results helps confirm the accuracy of the methods. Analytic
continuation of the QMC data is performed using the method
described in Ref.~\onlinecite{Wang09}.  The model has been previously
studied:\cite{Dopf92, Georges93, Zolfl98, Macridin05,
Weber08,Craco09,demedici09} At strong correlations and low dopings
the approximation reveals the ``Zhang-Rice singlet'' behavior with
doped holes residing on the oxygen but strongly
antiferromagnetically coupled to spins on the copper sites. As the
doping is increased or the effective correlation strength decreased,
the behavior reverts to moderately-correlated band behavior and a
Zhang-Rice picture becomes inappropriate.

From the calculation we obtain the electron spectral function
$A(\omega)={\rm Im} G(\omega)$ which we use to model the X-ray
absorption experiment. As noted in Ref.~\onlinecite{Hybertsen92},
the X-ray absorption cross section is not simply proportional to the
product of a matrix element and electron spectral function because
the hole in the oxygen $1s$ core state interacts with the excited
electron. We assume, following Ref.~\onlinecite{Hybertsen92} that
this effect provides a constant (Hartree) shift of the spectrum, and
in particular has a negligible effect on the lineshape and on
spectral weight. Therefore the x-ray absorption spectrum is given by
\begin{equation}
B(\omega)=C\cdot A_{oxy}(\omega-\omega_0)[1-f(\omega-\omega_0)],
\end{equation}
where $A_{oxy}(\omega)$ is the calculated electron spectral
function, projected onto the oxygen site, $[1-f(\omega)]$ is the
complement of the Fermi function, restricting the result to
unoccupied states, $C$ encodes the matrix element and $\omega_0$ is
the energy difference between the final and initial state, including
the effect of the excitonic interaction between the core hole and
the excited electron. In our calculations we set $\omega_0$ for 0.1
hole doping case such that the upper Hubbard band peaks at
$530.2eV$.\cite{Chen91} The $\omega_0$ for other doping values are
computed by adding a Hartree shift to the 0.1 hole-doped case, using
the Hamiltonian and parameters described by
Ref.~\onlinecite{Hybertsen92} with the calculated occupation
numbers.

\section{Results}\label{result}

\begin{figure}
    \centering
  (a)\includegraphics[width=6cm, angle=-90]{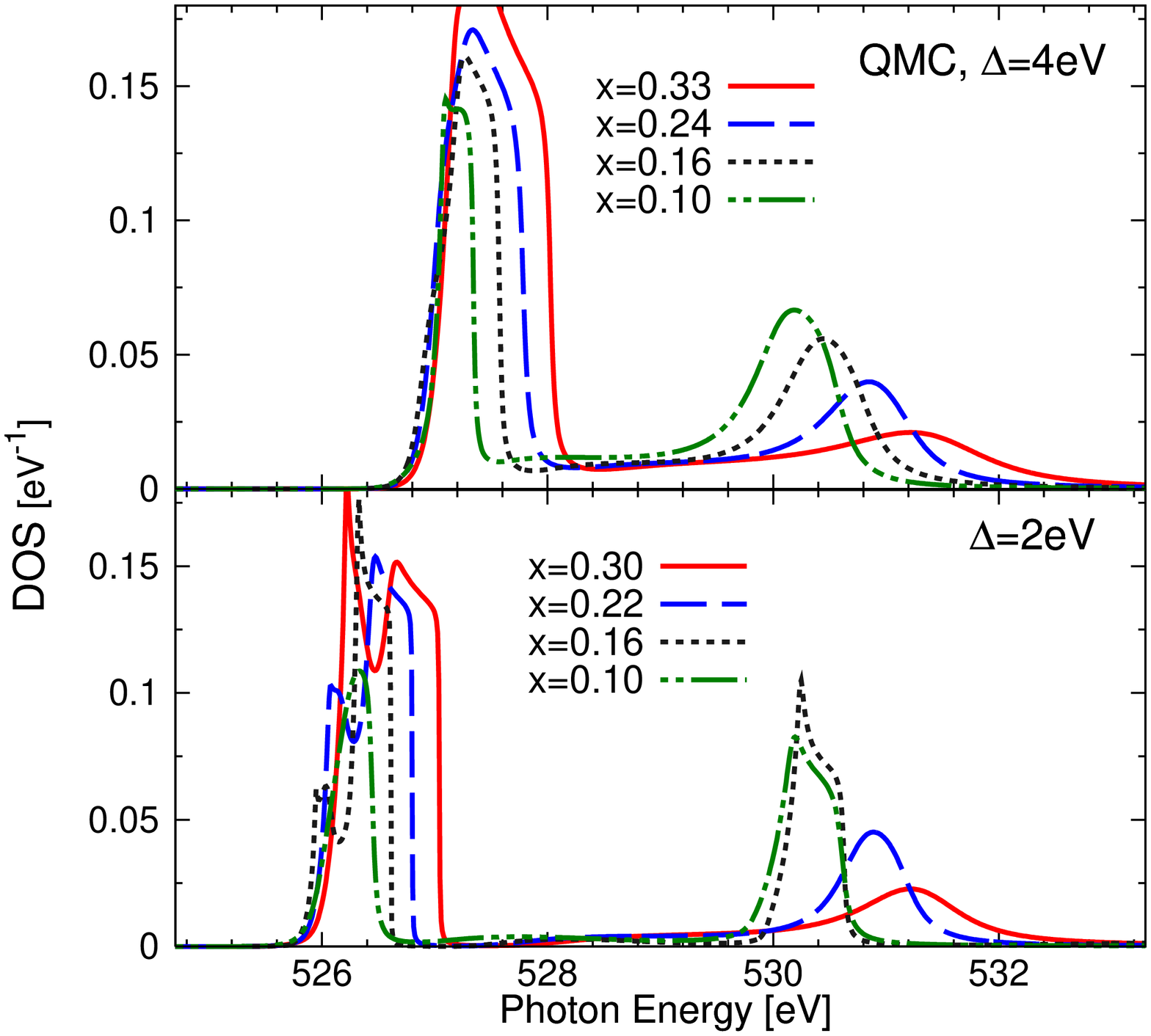}
  (b)\includegraphics[width=6cm, angle=-90]{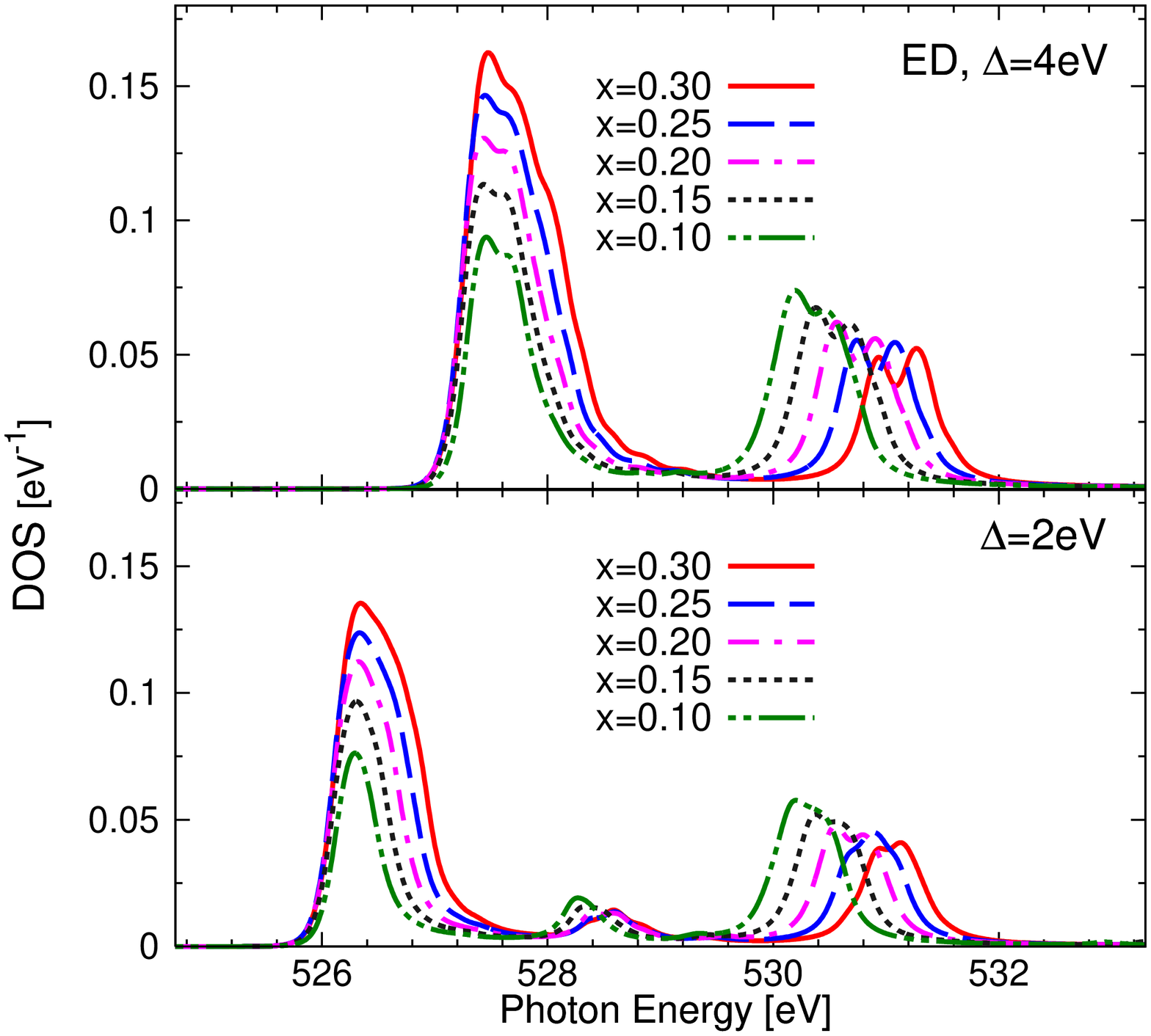}
    \caption{Unoccupied portion of three-band model many-body density of states,
    projected onto oxygen orbitals, calculated as described in the text for dopings $x$ indicated
    on figure using both analytical continuation of QMC [panels (a)] and
    ED [panels (b)] impurity solvers for two values of charge transfer gap
    $\Delta$; $\Delta=4eV$ (upper panels) corresponding to half filled paramagnetic metal
    and $\Delta=2eV$ (lower panels) corresponding to paramagnetic insulating phase at half filling.
    Parameters: $U=9eV$, $t_{pp}=0$, $t_{pd}=1.6eV$ and temperature $T=0.1eV$.}
    \label{tppzero}
\end{figure}

Fig.~\ref{tppzero} shows the doping dependence of the electron
spectral functions, calculated as described above and projected onto
the oxygen orbitals, for  $\Delta=2eV$ (insulator in undoped case)
and $4eV$ (metal in undoped case), in absence of $t_{pp}$.  Results
obtained from both analytic continuation of QMC and ED impurity
solvers are shown; the close correspondence between the results of
the two methods indicates that the solution of the model is
reliable.

Two features are evident in the calculated spectra: a higher energy
feature which corresponds to the upper Hubbard band (its weight is
small because this band is comprised mainly of $d$-states so it has
a small projection onto the oxygen states) and a lower energy
feature corresponding to the Zhang-Rice band. The difference in
energy between the two features is an estimate of the correlation
gap in the system. One sees that the gap size is about  $4eV$ in the
$\Delta=2eV$ (insulator in undoped case)  case, and around $2eV$ in
$\Delta=4eV$ (metal in undoped case) case, and we expect the gap
will further shrink if we go deeper in the metallic phase of the
paramagnetic phase diagram. Experiments  \cite{Chen91,Peets09} show
that the observed gap between the two peaks is  $\lesssim 2eV$,
which is more consistent with the $\Delta=4eV$ results, in agreement
with our previously published papers\cite{Comanac08,demedici09}
placing the cuprates on the paramagnetic metal side of the phase
diagram, so that the correlations are intermediate rather than
strong.

\begin{figure}
\vspace{.08in}
     \includegraphics[width=0.75\columnwidth, angle=-90]{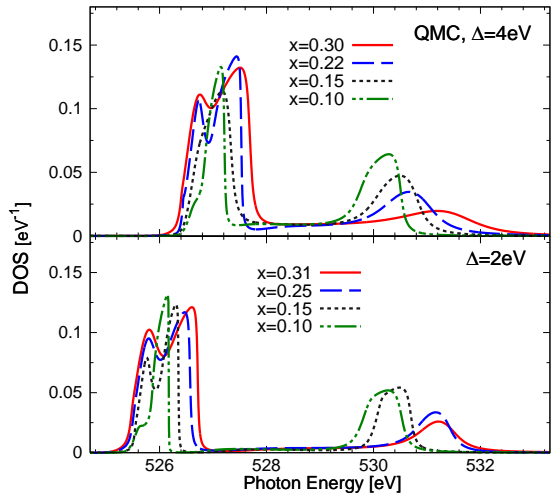}
    \caption{Calculated unoccupied oxygen density of states
    computed as a function of doping as
    described in the text for non-vanishing oxygen-oxygen hopping  $t_{pp}=1.1eV$
    by analytic continuation of QMC calculations.
    Other parameters: $U=9eV$, $t_{pd}=1.6eV$ and temperature $T=0.1eV$.}
    \label{tpp1p1}
\end{figure}

Fig.~\ref{tpp1p1} shows results for the case $t_{pp}=1.1eV$ obtained
from analytic continuation of the QMC data for the unoccupied
oxygen density of states.  The similarity of these curves to the
results shown in Fig~\ref{tppzero} indicates that  oxygen-oxygen
hopping does not have an important effect on the results.

\begin{figure}[t]
    \centering
    \includegraphics[width=0.75\columnwidth, angle=-90]{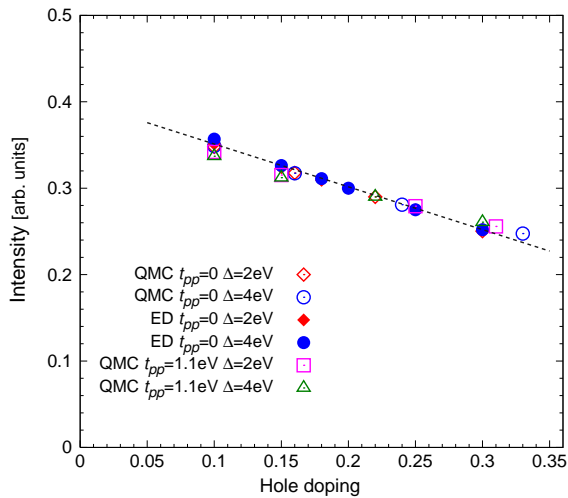}
    \caption{Calculated relative intensity of the peak associated with the upper Hubbard band
    of O $K$-edge x-ray absorption spectra v.s. doping.
    The results are normalized such that their values at 0.22 hole doping are the
    same. The dashed line is only guide to eyes.
    Parameters: $U=9eV$, $t_{pd}=1.6eV$, $T=0.1eV$, $t_{pp}$ and
    $\Delta=\varepsilon_p-\varepsilon_d$ as shown in the legends.}
    \label{UHB}
\end{figure}

We have integrated the area in the lower energy feature shown in
Figs.~\ref{tppzero}  and \ref{tpp1p1}, normalized the results to
the value at hole doping $0.22$  and have plotted the resulting
spectral weight as diamonds etc. in  Fig.~\ref{ZR}. For both values
of $\Delta$ the integrated spectral weight is found to increase
linearly with doping over a wide doping range. Differences between
the $\Delta=2eV$ case (strong correlation, doped charge-transfer
insulator)  and the $\Delta=4eV$ case (intermediate correlation,
charge-transfer metal) appear only at low dopings $x\lesssim 0.11$.
For completeness we present also in  Fig.~\ref{UHB}  the integrated
oxygen weight in the upper Hubbard band. We see that the spectral
weight of the Zhang-Rice band continue to increase at the overdoped
region, while the weight of the upper Hubbard band decreases. Both
the weight in the Zhang-Rice band and the weight in the upper
Hubbard band vary smoothly with doping for all parameters. The sharp
break in the data is not observed in the calculation for any choice
of parameters. We therefore conclude that the experimental paper
understated the significance of the results: the data indicate not
just a breakdown of the Zhang-Rice picture, but a failure of the
three-band model itself.

\section{Conclusion}\label{conclusion}

In this paper we have used single-site dynamical mean-field theory
to study the Emery three-band copper-oxide model related to the
cuprates. We have calculated the doping dependence of the intensity
of oxygen $K$-edge x-ray absorption spectra. At high doping, our
calculations does not reproduce the results of a recent experiment
(Ref.~\onlinecite{Peets09}). This implies a breakdown of
``Zhang-Rice singlet'' approximation and even a failure of the
entire three-band model picture.

We may speculate on the reason for the failure of the three-band
model in the overdoped regime. One possibility is that the
electronic structure changes in such a way that new degrees of
freedom (beyond the conventionally studied copper $d_{x^2-y^2}$ and
oxygen $p_\sigma$) states become important (for example other states
in the $d$-multiplet or apical oxygen states) providing a new
channel for adding doped holes. An alternative possibility might be
that for some reason the arguments relating the experimental
measurement to the oxygen density of states break down, or that
there is an unusual change in the matrix element. Such a change
would presumably be related to a change in the electronic structure.
A third possibility is that additional physics, such as a $U_{pp}$
on the oxygen states, begins to play a role; however it is not clear
why this would lead to a sudden change in doping dependence.  The
effect of additional orbitals is presently under investigation.

\section*{Acknowledgements}
XW and AJM are financially supported by NSF-DMR-0705847 and LdM by
Program ANR-09-RPDOC-019-01 and RTRA Triangle de la Physique. Part
of this research was conducted at the Center for Nanophase Materials
Sciences, which is sponsored at Oak Ridge National Laboratory by the
Division of Scientific User Facilities, U.S. Department of Energy.

\end{document}